\documentclass[sigplan,screen]{acmart}
\usepackage[utf8]{inputenc}
\usepackage{amsmath}
\usepackage{amsfonts}
\usepackage{url}
\usepackage{multicol}
\usepackage{caption}
\usepackage{subcaption}
\usepackage{makecell}
\usepackage{graphicx}
\usepackage{rotating}
\usepackage{listings}
\usepackage{xcolor}
\usepackage[inline]{enumitem}
\usepackage{multirow}

\setcopyright{none}

\newcommand{\LibName}{YASMIN}
\newcommand{\LibNameLong}{Yet Another Scheduling MIddleware for exploratioN}
\newcommand{\APIprefix}{}
\newcommand{\APIprefixName}{\texttt{yas\_}}

\newcommand{\ignore}[1]{}

\author{Benjamin Rouxel}
\email{b.rouxel@uva.nl}
\affiliation{%
  \institution{University of Amsterdam}
  \country{Netherlands}
}

\author{Sebastian Altmeyer}
\email{altmeyer@informatik.uni-augsburg.de}
\affiliation{%
  \institution{Augsburg University}
  \country{Germany}
}

\author{Clemens Grelck}
\email{c.grelck@uva.nl}
\affiliation{%
  \institution{University of Amsterdam}
  \country{Netherlands}
}

\begin{document}
\title{\LibName: a Real-time Middleware for COTS Heterogeneous Platforms}

\ccsdesc[500]{Computer systems organization~Embedded systems}
\ccsdesc[300]{Computer systems organization~Redundancy}
\ccsdesc{Computer systems organization~Robotics}
\ccsdesc[100]{Networks~Network reliability}

\keywords{Middleware, Real-Time Systems Deployment} 

\begin{abstract}
    
    

Commercial-Off-The-Shelf heterogeneous platforms provide immense computational power, but are difficult to program and to correctly use when real-time requirements come into play:
A sound configuration of the operating system scheduler is needed, and a suitable mapping of tasks to computing units must be determined. Flawed designs may lead a sub-optimal system configurations and thus to wasted resources, or even to deadline misses and failures. 

We propose \textit{\LibName}, a middleware to schedule end-user applications with real-time requirements in user space and on behalf of the operating system.
\textit{\LibName} provides an easy-to-use programming interface and portability.
It treats heterogeneity on COTS heterogeneous embedded platforms as a first-class citizen: It supports multiple functionally equivalent task implementations with distinct extra-functional behaviour. This enables the system designer to quickly explore different scheduling policies and task-to-core mappings, and thus, to improve overall system performance. In this paper, we present the design and implementation of \textit{\LibName} and provide an analysis of the scheduling overhead on an Odroid-XU4  platform. 
Last but not least, we demonstrate the merits of \textit{\LibName} on an industrial use-case involving a Search \& Rescue drone.
\end{abstract}

\maketitle

\section{\label{sec:intro}Introduction}

Commercial-off-the-shelf (COTS) heterogeneous parallel platforms are very popular as they offer (in this hardware segment) unprecedented computational power at low cost.
They typically combine a potentially heterogeneous multi-core CPU (e.g.~ARM big.LITTLE) with a powerful GPU and, possibly, various additional hardware accelerators \cite{nvidiatx2}.
A large segment of embedded computing must meet real-time constraints while not being safety-critical, e.g.
Internet-of-Things (IoT) \cite{vermesan2011internet} or edge computing \cite{tian2019real}, including a range of cyber-physical systems (CPS) \cite{calvaresi2017challenge}. Supporting real-time applications targeting such platforms is insanely complex as all required analyses need to be adapted for each computing units type. Moreover, heterogeneous parallel architectures immediately create a complex scheduling and mapping problem of application tasks to execution units, with highly different timing and energy properties. Furthermore, guaranteeing real-time properties on such systems after deploying these applications is often a nightmare as the execution environment is mostly constrained by vendor-provided (or vendor-adapted) operating systems (OS), which most often lack support for real-time techniques exhibited by the research community.

To enforce timing properties when deploying real-time applications, designers have the choice to use a modified kernel, e.g.~\textit{Litmus\^{}RT} \cite{calandrino2006litmus}, or real-time patches for general-purpose OS \cite{balsini2014sched}. However, these solutions preclude using specific hardware drivers used in vendor-specific OS setups and, therefore, 
are of limited effectivity in practice.
In fact, COTS embedded heterogeneous platforms are mostly bound to specific OS version, e.g.~the Apalis TK1 \cite{toradextk1} board can only run a Linux v3.10 kernel.
Likewise, proprietary device drivers, mainly for the embedded GPU and further accelerators, set tight limits to change or modify the OS, not to mention that kernel modifications are cumbersome, difficult, error-prone and non-portable.

We propose a novel middleware: \LibName{} (\LibNameLong) that facilitates the deployment of real-time applications on heterogeneous COTS platforms running atop a COTS OS. Compare to previous real-time-related attempt, e.g. \cite{akesson2020empirical}, \LibName{} is, to the best of our knowledge, the first middleware to embrace heterogeneity as a central design concern.

A recent survey by Akesson et al. \cite{akesson2020empirical} shows that industrial practitioners are very keen on using COTS OS in conjunction with libraries to deploy real-time systems. Hence, \LibName{} fits industrial needs: 
\begin{itemize}
    \item \textit{Customisation}: \LibName{} is highly customisable through clear separation of concerns (mapping, scheduling, priority ordering \ldots). Thus, adding a new state-of-the-art technique to \LibName{} is much simpler and faster than adding it into an OS kernel such as Linux.
    
    \item \textit{Adaptability}: Through multiple implementations \LibName{} permits users to change the behaviour of the application at run-time to cope with constantly evolving environmental constraints, such as the detection of a fault, a cyber-attack, a drop of the available energy.
    \item \textit{Maintainability}: \LibName{} is not dependent on any specific OS or OS version. Hence, upgrading a system to benefit from security patches or bug fixes is considerably easier using \LibName{} than with a deployment environment bound to a specific kernel, e.g.~\cite{aasberg2012exsched}.
    \item \textit{Portability}: \LibName{} requires to run atop of a POSIX-compliant OS, and is not bound to a specific platform. Therefore, executing the application compiled with \LibName{} on different platforms merely requires recompilation.
    \item \textit{Compatibility}: When a specific platform has no RTOS support, e.g.~due to vendor-specific drivers, \LibName{} provides more timing guarantees than what a vanilla OS has on offer, e.g.~only soft-real-time applications can be enforced on a vanilla Linux with few configurations \cite{rouxel2020prego}.
    \item \textit{Flexibility}: Using \LibName{} with different workload packages supports different configurations for each package, such as different task model, scheduling strategies, etc. This property is shown by the aforementioned survey \cite{akesson2020empirical} to be a prerequisite to deploy industrial systems.
    \item \textit{Design Exploration}: Deciding which scheduling policy is the best for a system is rarely trivial. \LibName{} offers multiple scheduling options, which can be switched at compile time. Hence, RT-experts and non-experts alike can explore the scheduling design space to select the best performing technique.
\end{itemize}

\LibName{} is part of a more comprehensive endeavour to facilitate rapid prototyping and deployment of non-safety-critical real-time applications targeting heterogeneous parallel COTS systems.
Application components, their functional interplay, timing properties and requirements can be specified in a high-level coordination DSL~\cite{roeder2020towards}.
Following a mostly automated generative approach~\cite{rouxel2020prego}, our compiler tool chain turns a high-level description of an application into C code ready for binary code generation by a target-specific C compiler.
The whole tool chain, including \LibName, is available under a GPLv3 licence \cite{splorer}. 

The remainder of this paper is organised as follows:
In Section~\ref{sec:bg} we discuss fundamental assumptions such as the underlying task model.
In Section~\ref{sec:implem} we present design and implementation of \LibName{} and elaborate on the various options and design choices we support.
We empirically validate \LibName{} in Section~\ref{sec:exp}, and  we apply it on an industrial use-case in Section~\ref{sec:use-case}.
We review related work in Section~\ref{sec:rw} and draw conclusions in Section~\ref{sec:concl}.

\section{\label{sec:bg}Application model}

We consider non-safety-critical real-time systems composed of a set of sporadic, periodic tasks where each task represents an indivisible (or atomic) feature of the end-user application. The minimal time interval $T$ (or period) separating two consecutive task activations must be provided to our scheduler. In addition, we allow additional aperiodic tasks managed by the end-user as no regular pattern can be given to the scheduler. Real-time tasks must complete their execution before a deadline $D$ relative to the period. We support the three main deadline schemes: implicit ($D=T$), constrained ($D\leq T$) and arbitrary to the period.

To embrace heterogeneity, we adopt recent task models representing each task with a set of versions \cite{roeder2020towards}, or variants \cite{houssam2020hpc}.
All versions of a single task are functionally equivalent, and expose the same interface (i.e.~inputs, outputs), but each one has its own distinct non-functional behaviour, i.e. worst-case execution time (WCET), energy consumption.

The immediate motivation for multi-version tasks lies in the scheduling and mapping complexity with heterogeneous platforms, where it is commonly not a-priori decidable which tasks should exclusively run on the CPU and which should exclusively run on one (or more) of the various accelerators. Consider an application with at 2 tasks ${A, B}$, and each task has 2 versions running: \begin{enumerate*}
    \item 100\% on CPU,
    \item 1\% CPU and 99\% GPU.
\end{enumerate*}
These 2 tasks are independent and have the same period. Hence, they could potentially run in parallel. On the target platform, however, only 1 GPU is available. Therefore, both versions of $A$ and $B$ targeting the GPU cannot execute in parallel. However, the presence of different versions allows us to run the GPU version of $A$ at the same time as the CPU version of $B$, or vice versa, as long as the CPU version doesn't overrun the sequential execution of the 2 GPU versions. We empirically demonstrated in \cite{roeder2020towards} that deciding which version to execute at each task instance is not straightforward.
This question is rather part of the scheduling problem, and it is common that depending on global circumstances and objectives, the same task may sometimes preferably be executed on the CPU and in other cases on the GPU, see \cite{roeder2020towards} for details.

The versatility of multi-version tasks goes even beyond the above. The computing unit heterogeneity may exhibit different ISAs per core, or they can likewise provide task implementations particularly optimised for execution on a specific HW unit, even in the presence of generic ISA compatibility.
Furthermore, application designers could easily play with implementation variants that expose different non-functional behaviour (e.g. energy, time, security) and let \LibName{} automatically select the best suited one under concrete context and objectives.

\LibName{} further supports tasks grouped into graphs with precedence constraints, thus forming a so-called Directed Acyclic task-Graphs (DAG). Other graph-based task models, such as Synchronous DataFlow (SDF) \cite{lee1987static}, must \textit{a-priori} be transformed (or expanded for SDF) to comply with a DAG task model. 
As in most graph-based task models, \LibName{} supports activation patterns and relative deadlines described at the graph level: The whole graph is considered sporadic or periodic. 

\section{\label{sec:implem}\LibName~ Design \& Implementation}

We designed \LibName{} as a library to be compiled individually and linked to the end-user program. 
\LibName{} is highly modular and allows \begin{enumerate*}
    \item the use of various scheduling policies and, 
    \item easy switching between them at compile time using a configuration header file. 
\end{enumerate*}

We implemented \LibName{} in structured C-code following real-time and MISRA-C 2012\footnote{We checked for MISRA-C compliance using the trial version of PC Lint Plus \cite{pclint}} coding guidelines to enable the use of WCET analysis tools, such as AbsInt's aiT \cite{ferdinand2004ait} or Heptane \cite{hardy2017heptane}.
We systematically refrain from using dynamic memory allocation, and loops are statically bounded. To accomplish this, we make use of C-header configurations to define constants used throughout the library, e.g.~the number of threads or the number of tasks. 

\LibName{} is compatible with any POSIX compliant OS. However, we also rely on the \textit{pthread\_set\_affinity\_np} non-POSIX function that binds a thread to a specific core. Similar requirements can be found in previous works \cite{mollison2013bringing, saranya2014implementation}.

\subsection{\label{ssec:implem:gen}\LibName{} API}

The library is configured at compile time using a configuration file. In this file, pre-processor definitions set, among others, the type of scheduling, the type of mapping and the priority assignment. 
Each different scheduling strategy requires different mandatory information to perform adequately, but we kept a uniform interface for all scheduling configurations.
The configuration is applied to the whole compiled binary, only one scheduling policy is allowed at a time. In order to switch to another policy, the application must be recompiled with new parameters.

\begin{table}[ht]
    \centering
	\caption{Full API of \LibName}
	\label{tbl:implem:api}
	\begin{tabular}{|l|l|}
		\hline
	    \makecell[l]{struct TData \{\\
	        \quad char *name,\\
	        \quad u64 period,\\
	        \quad u64 deadline,\\
	        \quad u16 virt\_core\_id,\\
	        \quad u64 release\_offset\}} & \makecell[l]{Structure to describe a task. \\Some fields are optional \\depending on the configured \\scheduling policy.} \\
		\hline
        void \APIprefix init(void) & \makecell[l]{Initialise \LibName.}\\
		\hline
		void \APIprefix cleanup(void) & \makecell[l]{Wait for all worker threads\\ to finish and close.}\\
		\hline
		
        bool \APIprefix start(void) & \makecell[l]{Start to execute the tasks.}\\
		\hline
		void \APIprefix stop(void) & \makecell[l]{Stop pushing new tasks into \\the ready queue. All tasks already\\pushed will be executed.}\\
		\hline
		
		\makecell[l]{TID \APIprefix task\_decl(\\
    	\quad TData *d)} & \makecell[l]{Declare a task to the scheduler. }\\
		\hline
		\makecell[l]{void \APIprefix task\_activate(\\
		\quad  TID t)} & \makecell[l]{Activate a non-recurring task \\for immediate schedule.}\\
		\hline
		
		\makecell[l]{VID \APIprefix version\_decl(\\
		\quad  TID t,\\
		\quad FuncPtr f,\\
		\quad void *f\_static\_args,\\
		\quad VSelect props)} & \makecell[l]{Add a version to the task \\with user specific properties.}\\
		\hline
		
		\makecell[l]{HID hwaccel\_decl(\\
		\quad char *name)} & Declare a hardware accelerator\\
		\hline
		\makecell[l]{void hwaccel\_use(\\
		\quad TID t, \\
		\quad VID v, \\
		\quad HID a)} & \makecell[l]{Declare a hardware accelerator \\used by a task version.}\\ 
		\hline
		
		\makecell[l]{channel\_decl(\\
		\quad CID, \\
		\quad datatype,\\
		\quad size)} & \makecell[l]{Macro to declare a channel \\of type $type$ identified by $CID$ \\ containing $size$ items \\of type $datatype$. }\\
		\hline
		\makecell[l]{channel\_connect(\\
		\quad TID src,\\
		\quad TID dst,\\
		\quad CID)} & \makecell[l]{Macro to connect a source \\and a destination task \\using the specified channel \\identified by $CID$.} \\
		\hline
		\makecell[l]{\APIprefix channel\_push(\\
		 \quad CID,\\
	     \quad datatype d)}& \makecell[l]{Macro to push a value of \\type $datatype$ in the FIFO \\identified by $CID$. \\To be used in user function body.}\\
	     \hline
	     \makecell[l]{\APIprefix channel\_pop(\\
	     \quad  CID,\\
	     \quad datatype *d)}& \makecell[l]{Macro to pop a value of \\type $datatype$ in the FIFO \\identified by $CID$. \\To be used in user function body.}\\
	    \hline
		
		\end{tabular}
		
\end{table}

Table~\ref{tbl:implem:api} presents the API of \LibName{}. All functions are prefixed with \APIprefixName{}, which we left out in the paper for conciseness. This API is common to all scheduling strategies, allowing for an easy switch at compile time without modifications of the user code if all information are provided.

The end-user program must first call the \textit{\APIprefix init} function that initialises different structures of our library. Then, the user must declare the various tasks using \textit{\APIprefix task\_decl} and their associated versions with \textit{\APIprefix version\_decl}. 

\LibName{} supports graph-based tasks. We provide a mechanism to declare and manage FIFO channels required between causally dependent tasks within a graph. The pre-processor macro \textit{\APIprefix channel\_decl} defines the FIFO channel buffer. Connecting two tasks to use this channel is done with \textit{\APIprefix channel\_connect}. The channel can be accessed from within user tasks with the \textit{channel\_push} and \textit{channel\_pop} functions. 

Hardware accelerators can be declared with \textit{hwaccel\_decl} and linked to a task version with \textit{\APIprefix hwaccel\_use}. The scheduler is therefore aware of accelerator usage, and can apply smart strategy to select a version at runtime according to some criteria, see Section \ref{ssec:implem:hetero}.

At this stage no user code has yet been executed, and no scheduling has been performed. It is after the call to \textit{\APIprefix start} that the scheduler starts to run the application. Calling the \textit{\APIprefix stop} function stops the scheduler. Then, either the main program performs the finalisation of the application with \textit{\APIprefix cleanup}, or the schedule can be resumed with a new call to \textit{\APIprefix start}. It is only possible to alter the task set while the schedule is not running, hence enabling multi-mode scheduling \cite{goossens2019acceptor}. Functions to alter the task set are for conciseness removed from the following API tables.

\subsection{\label{ssec:implem:hetero}Heterogeneity \& Multi-Version}

With embedded platforms hardware accelerators are usually a scarce resource, i.e.~there is typically only 1 GPU. If multiple tasks need to access an accelerator then they might need to wait for the resource to become available. To avoid this form of congestion we introduce multi-version tasks. A task may have 1 implementation targeting the GPU, 1 using some other accelerator, and yet another targeting the CPU. Because accelerator usage is declared to our scheduler using the API call \textit{hwaccel\_use}, it can detect that the targeted accelerator is busy, and that it is preferable to use another task version targeting a free one. 

Should our scheduler not be able to determine a matching version where all hardware resources are available, and if the current task has a higher priority than the one currently using the targeted resource, we apply a Priority Inheritance Protocol (PIP) \cite{rajkumar2012synchronization} and reschedule the task.

Going further with versions, we provide multiple configuration options to automatically select the version to use for the current job. At the time of writing it is possible to configure the version selection depending on \begin{enumerate*}
    \item the current energy capacity of the platform,
    \item depending on an energy/time trade-off,
    \item depending on the current execution mode\footnote{For example, multi-security mode where different implementations of an encryption algorithm can be switched at runtime by changing the mode of execution.},
    \item depending on a bit mask permission, or
    \item with a call to a user-defined function.
\end{enumerate*}
The method to use is specified in the configuration header file, thus, only one method is effectively used at runtime, but switching is possible at compile time. 

Each of these selection options requires different information from the user. They are provided when declaring a version using \textit{version\_decl} through the \textit{VSelect props} argument. The type of this argument is a structure morphed to cope with the selected method. For example, if the method to select the version is based on the energy then the structure includes two fields to provide the energy budget of the task, and a user function to request the platform-dependent battery status. An example is given in Section~\ref{ssec:implem:example}.

\textbf{Limitation: }
Practically, task versions targeting a specific hardware accelerator start on a CPU core before they move the main workload to the accelerator, and eventually complete their execution back on a CPU core. For the time being, we consider the accelerator busy from the beginning of the initial CPU part to the end of the final CPU part. In the near future we plan to add an asynchronous mechanism, where CPU cores can be used by tasks while the accelerator-bound task actually runs on the accelerator.

\subsection{\label{ssec:implem:part-on-line}Partitioned \& Global On-line Scheduling}

We rely on the concept of shielded processors, as described in \cite{saranya2014implementation,brosky2003shielded}. The idea is to reserve cores to only execute real-time (RT) tasks in order to minimise interference with system tasks. On each of the reserved cores we spawn 1 thread, so-called \emph{worker threads} or \textit{virtual CPUs}, which serve as containers for the execution of the user RT tasks.

An on-line scheduler must activate tasks following their arrival time (period), decide which version of the task to execute, and dispatch tasks to a worker thread. Two modes are available: \begin{enumerate*}
    \item \textit{Global} when all tasks can be executed on any virtual CPU, and
    \item \textit{Partitioned} when all tasks have a predefined target virtual CPU.
\end{enumerate*}
The selection between the two modes is done at compile time through the configuration header file. Hence, only one of the two options is effectively compiled into the resulting binary. Switching between global and partitioned scheduling requires the modification of a single macro definition and a recompilation.

\LibName{} supports static and dynamic priority assignments following task periods (rate monotonic), deadlines (deadline monotonic, earliest deadline first) or any statically user-defined priorities.

Specifically with graph-based tasks, only the root nodes need to have a period attached. Subsequent nodes are automatically activated by the scheduler, once all required incoming data are present in their input channels. 

\begin{figure}[ht]
    \begin{subfigure}{0.32\textwidth}
        \centering
        \includegraphics[width=1\textwidth]{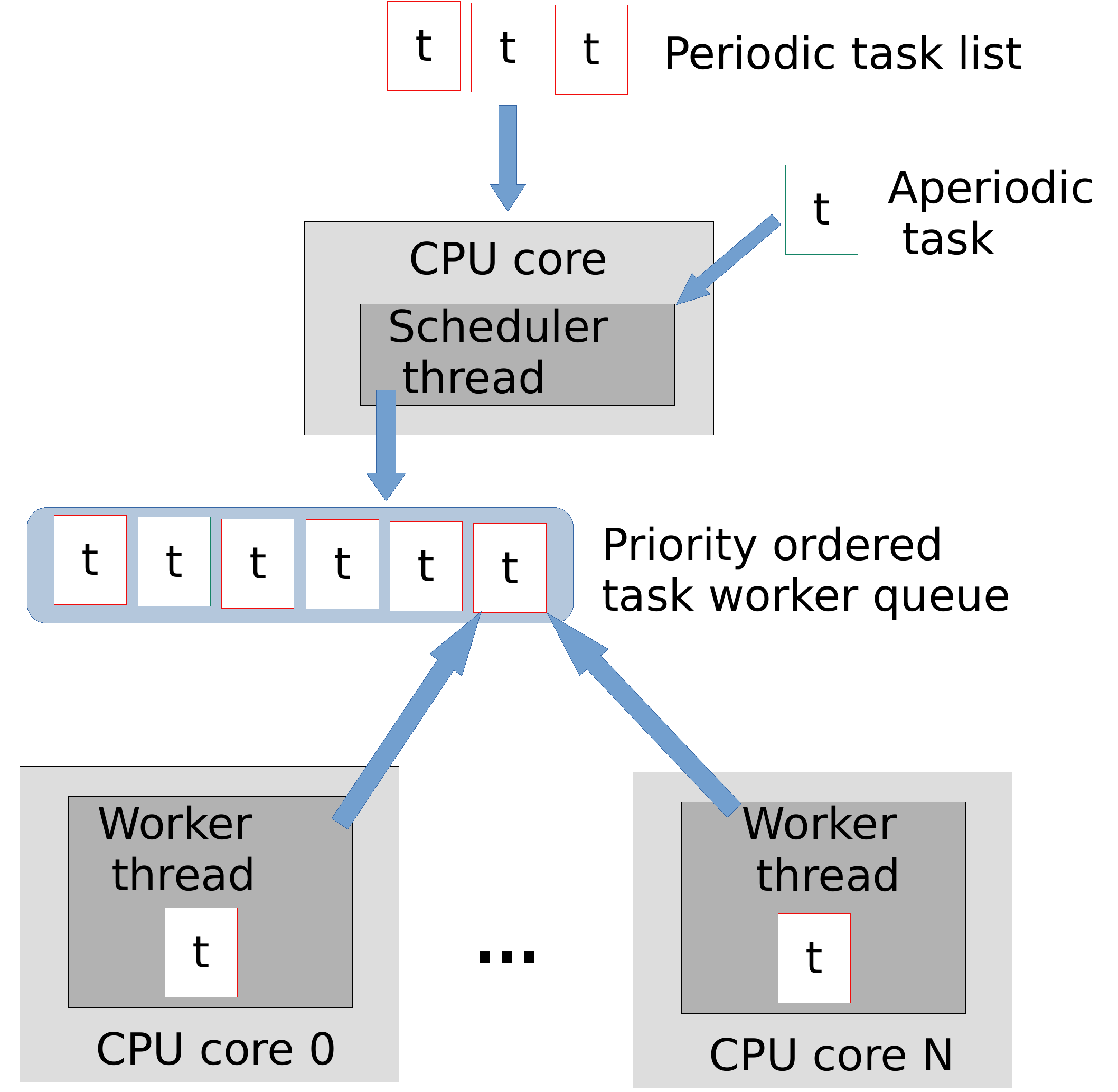}
    	\caption{Global on-line scheduling strategy. 
    	The ready task queue is shared among worker thread.}
    	\label{fig:implem:glob-on-line}
    \end{subfigure}
	\begin{subfigure}{0.32\textwidth}
        \centering
        \includegraphics[width=1\textwidth]{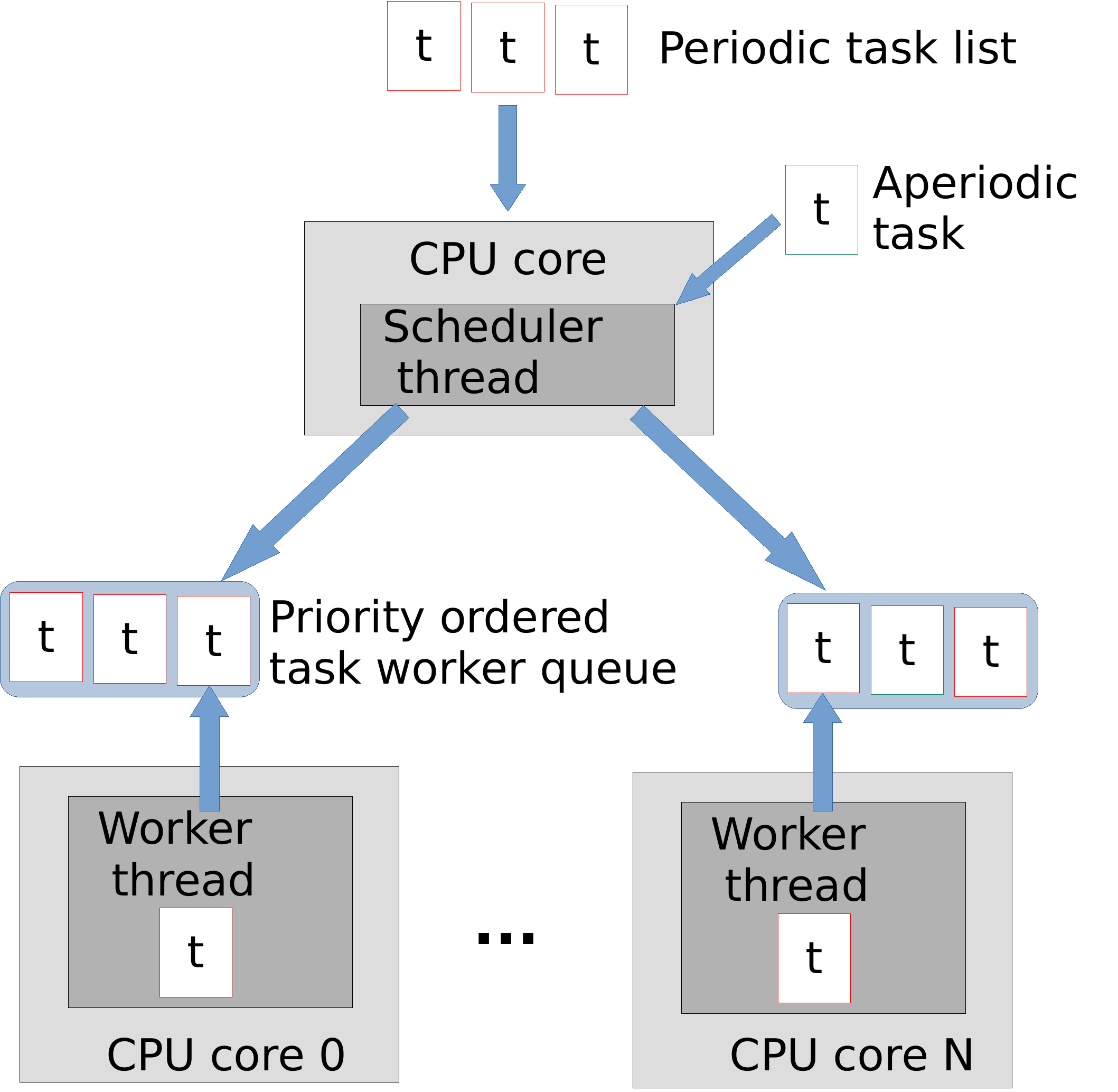}
    	\caption{Partitioned on-line scheduling strategy. 
    	A scheduler thread pinned to another core feeds each worker thread ready task queue.}
    	\label{fig:implem:part-on-line}
    \end{subfigure}
    \begin{subfigure}{0.33\textwidth}
        \centering
        \includegraphics[width=1\textwidth]{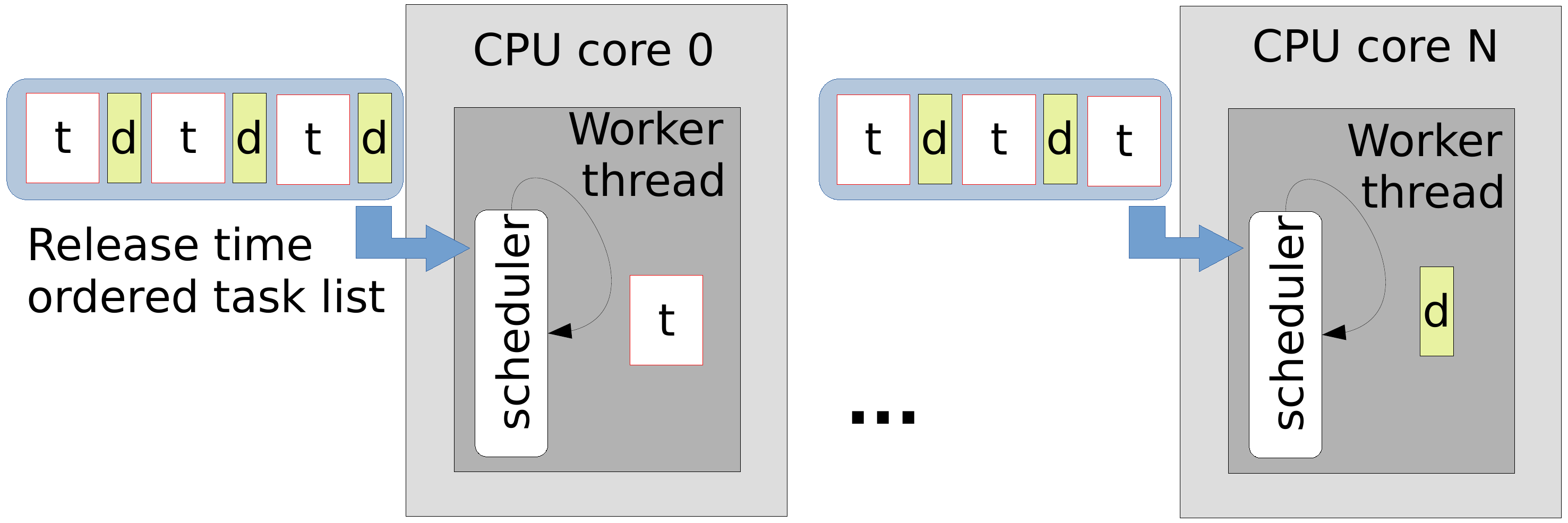}
    	\caption{On-line dispatcher with an off-line scheduling strategy. 
    	Each worker thread is pinned to a core, and a scheduling loop iterates on its ready task queue.}
    	\label{fig:implem:off-line}
	\end{subfigure}
    \caption{Overall architecture for each scheduling class.}
    \label{fig:implem:overview}
\end{figure}

Figures~\ref{fig:implem:glob-on-line} and~\ref{fig:implem:part-on-line} respectively illustrate our overall architecture for global and partitioned scheduling strategies. In both mode, each worker thread is pinned to a specific core.
With global scheduling all worker threads share a common ready queue, whereas with partitioned scheduling each worker thread has its own ready queue.

In either case, global or partitioned, the ready queue is filled by a separate scheduler thread that is likewise pinned to its private core. Unlike in \cite{saranya2014implementation}, who also uses an external scheduler thread, we do not constantly check for new tasks to activate. Instead, we only periodically check for new tasks to schedule, i.e.~between two activations the scheduler thread \textit{waits}. 
The period of the scheduler thread is determined using the greatest common divisor of all the declared task periods.

Using a separate scheduler thread, that executes on its private core, decreases parallelism, as one core less is available to execute user RT tasks, but it increases predictability by minimising interference with user RT tasks. For example with Linux, a core pinned scheduler task from the kernel periodically preempts the running thread to check for higher priority tasks to execute. To fulfill real-time requirements, this blocking mechanism must be accounted for in the worst-case response time of user RT tasks. However, in practice, it is very difficult to estimate this blocking time spent in the kernel scheduler. Using a separate scheduler thread to check for higher priority tasks avoids such blocking, and still allows preemption.

In addition, it is possible to configure the Linux kernel to prevent the aforementioned periodic scheduler task: a value of $-1$ needs to be written in the virtual file \textit{/proc/sys/kernel/sched\_rt\_runtime\_us}. We refer, the interested reader on how to increase user control over time in Linux to \cite{rouxel2020prego}.

\textbf{Limitation:} 
%
We do not support job migration. A job (task instance) spawned on a virtual CPU cannot be migrated to another one. However, we do support task migration: job $i$ of some task may run on one virtual CPU while job $i+1$ runs on a different virtual CPU.

\subsection{\label{ssec:implem:off-line}Off-line Scheduling}

Unlike any similar middleware we found in literature, \LibName{} also natively supports \textit{off-line} computed schedules. 
An \textit{off-line} schedule is computed before executing the application using the timing properties of the task set. 
In our run-time implementation an \textit{on-line} dispatcher dispatches tasks at the predefined time following a given time table and a given mapping.

Figure~\ref{fig:implem:off-line} presents the overall architecture for the off-line scheduling strategy. Each worker thread is pinned to a specific core and has access to a predefined sequence of RT tasks ordered by increasing release time. Upon creation each worker thread starts executing a control loop running the RT task in order. To respect the release time of each task (computed off-line), special delay slots are added in between RT tasks that make the worker threads \textit{wait} for a pre-computed duration. 

If the static scheduler is aware of multi-version tasks, the version can be pre-selected off-line. This has the advantage of reducing the size of the resulting binary size as it only needs to embed the actually required task versions.

\textbf{Limitation: }
We consider heterogeneous resource management 
to be handled by the off-line scheduling step. A task can, hence, target an accelerator without requesting access to the on-line dispatcher.

\subsection{\label{ssec:implem:other}Further Implementation Aspects}

This section describes other design issues we encountered and how we addressed them in \LibName.


\textbf{Accessing time: }
We access time using the POSIX primitive \textit{clock\_gettime} where the given clock can be set using the configuration file. As default, \textit{CLOCK\_MONOTONIC} is employed. It gives a monotonically increasing clock with nanoseconds precision. The POSIX standard does not specify what the time $0$ means. In Linux time $0$ corresponds to system boot time. Our library stores the time at which the schedule is started using API call \textit{\APIprefix start}. Afterwards, all timing information is computed using this initial starting time.

\textbf{Pre-emption: }
\LibName{} supports pre-emption with on-line scheduling policies only. 
Upon sorting, similar to \cite{mollison2013bringing}, the scheduler thread sends a signal (PREEMPTION\_SIGNAL), using the \textit{pthread\_kill}~POSIX primitive, to each worker thread executing tasks with a lower priority than that of the head of the ready queue. This signal is caught by the thread which looks in the ready queue for a higher priority task. If a higher priority task is found, a context-switch is operated. Upon completion, the process of finding a higher priority task is repeated until the initial preempted task becomes the highest priority one, and the context is switched back to it. 

\textbf{Context switching: }
Similar to \cite{mollison2013bringing} we use an architecture-dependent \textit{swapcontext} function (in assembly code), which is called when switching execution context upon pre-emption. 
We draw inspiration from the GLibC swapcontext implementation, but leave out extra syscalls. As of writing, our swapcontext implementation is available for ARM 32/64 bits as well as X86-64 architectures.

\textbf{Locking: }
Internally we implement synchronisation primitives, i.e.~mutex locks and barriers, in two different manners: A first implementation uses the POSIX API implemented in the kernel and GLibC. A second implementation relies on lock-free algorithms from \cite{mellor1991algorithms}. It is possible to select one of the two options at compile time using the configuration file. We believe that lock-free algorithms form a superior choice for static WCET analysis \cite{mellor1991algorithms}, but spinlocks exhibit higher energy consumption. On the other hand, it is hard to analyse kernel and GLibC calls, but this soultion offers better energy performance at the cost of predictability due to the kernel replacing the worker thread by an internal idle task. Selecting one or the other option depends on user preferences regarding predictability and energy conservation.

\textbf{Waiting:}
With similar consideration in mind, we provide the option to configure the waiting strategy in two ways: 
\begin{enumerate*}
\item \textit{sleep} (default): calls some kernel code, which is hardly timing-analysable, 
\item \textit{spinlock}: enable a more precise overhead analysis at the cost of potential energy waste.
\end{enumerate*}

\textbf{Protecting against page fault: }
Similar to \cite{mollison2013bringing} we lock our library code in memory using the POSIX primitive \textit{mlockall}. This prevents swapping out the code of our library.

\textbf{Interrupts: }
We set the kernel to use \textit{threadirq}, and we shield the processor using \textit{isolcpu}. Hardware interrupt handlers are composed of two parts, a top and a bottom part. We cannot do much about the top part that usually is pinned to a specific core. For the bottom part, if they are not pinned to a specific core, then the same configuration as for software interrupts applies. If they are specific to a core, and this core runs a worker thread, or the scheduler thread, then their schedule is left to the underlying OS. Care must therefore be taken to ensure that the priority of our worker threads, and/or scheduler threads allows these bottom part interrupt handlers to execute.



\subsection{\label{ssec:implem:example}Example}
The following two listings~\ref{lst:implem:hsnippet} and~\ref{lst:implem:csnippet} show an example of four tasks. The four tasks represent a diamond graph where a fork is connected to two other tasks before joining to a join task. Data are exchanged using FIFO channels. The task \textit{left} has two versions, one using a specific hardware accelerator, and the other not. \LibName{} is configured to select the version according to the current energy capacity of the platform.

\begin{lstlisting}[language=C,caption={Essential configuration example, must be in a \textit{config.h} file},label=lst:implem:hsnippet]
#include "yasmin_constants.h"
/*there 1 periodic task: the fork task*/
#define PERIODIC_TASK_SIZE 1 
/*all other tasks are activated depending on the presence of input data*/
#define NONPERIODIC_TASK_SIZE 3  
/*there are 4 FIFO channels connecting tasks*/
#define CHANNEL_SIZE 4  
/*At most 2 versions are used*/
#define VERSION_MAX_SIZE 2  
/*adapt the structure and code to select versions of task based on remaining energy.*/
#define VERSION_SELECTION ENERGY 
/*One hardware accelerator is used*/
#define HWACCEL_SIZE 1  
/*the example uses a global on-line scheduler*/
#define MAPPING_SCHEME  GLOBAL  
/*priority are given using EDF*/
#define PRIORITY_ASSIGNMENT EDF  
/*2 worker threads will be used*/
#define THREADS_SIZE 2 
\end{lstlisting}
\begin{lstlisting}[language=C,caption=C code example using \LibName{} common API with user-defined priority,label=lst:implem:csnippet]
struct token { int value ; }
/*declare a dependency without data exchange*/
channel_decl(fl, char, 0);
/*declare dependencies with data exchange*/
channel_decl(fr, struct token, 1);
channel_decl(rj, int, 2);
channel_decl(lj, int, 1);

void fork(void *arg) {          
  struct token; token.value = 2;              
  channel_push(fr,token)        
}                               
void right(void *arg) {                                
  struct rec_token;       
  channel_pop(fr, &rec_token);
  channel_push(rj, rec_token.value);
  channel_push(rj, rec_token.value*2);
}
void join(void *arg) {          
   int rec_data;                
   channel_pop(rj, &rec_data);  
   channel_pop(rj, &rec_data);  
   channel_pop(lj, &rec_data);  
}                               
void left_v1(void *arg) {                                
  int *a = (int*) arg;                           
  channel_push(l, *a);    
}
void left_v2(void *unsued) {
  int val = get_val_from_specific_accel();
  channel_push(l, val);
}
/*User defined function to get the battery status*/
static void current_battery_level() { return .... ; }
void main(int argc, char **argv) {
  TData f, j, r, l;        TID fid, jid, rid, lid;
  VID lv1id, lv2id;     HID aid;
  /*Due to the given configuration, the required information to select a version is energy budget*/
  VSelect lv1_select, lv2_select; 

  init(); // initialise @\LibName{}@

  f.name = "fork"; f.period = 250;
  //initialise other tasks
  l.name = "left";
  lv1_select.energy_budget = 5;   
  lv2_select.energy_budget = 12;
  lv1_select.get_battery_status = 
    lv2_select.get_battery_status = current_battery_level;

  fid = task_decl(&f, fork, NULL);
  //declare other tasks
  lid = task_decl(&l);
  lv1id = version_decl(lid, left_v1, lv1_select);
  lv2id = version_decl(lid, left_v2, lv2_select);

  aid = hwaccel_decl("quantum_rand_num_generator");
  hwaccel_use(lid, lv2id, aid);

  channel_connect(fid, rid, fr);      
  channel_connect(rid, jid, rj);
  //declare other channel connections

  start(); // Start the schedule
  //wait for some event
  stop(); // Stop the schedule
  cleanup(); // Cleanup before exiting
  return 0;
}  
\end{lstlisting}
\section{\label{sec:exp}Evaluation}

We empirically evaluate the overhead and latency introduced by \LibName{} against various state-of-the-art task management. 

We target the embedded heterogeneous COTS platform Odroid-XU4 \cite{odroidxu4}, as it provides multiple heterogeneous cores and allows us to run a RTOS with the support of the PREEMPT\_RT patch set for Linux. The Odroid-XU4 platform includes an ARM big.LITTLE octa-core CPU and a Mali GPU. The CPU is split into two clusters: the LITTLE cluster contains four energy-efficient but computationally less powerful ARM Cortex-A9 cores while the big cluster embeds four computationally powerful but energy-greedier ARM Cortex-A57 cores.
We configure the OS following our guideline to tame Linux and minimise interference between OS and application code \cite{rouxel2020prego}.
All code is compiled with GCC 4.9 without optimisation ($-O0$), as is common to perform WCET analysis for real-time systems \cite{wilhelm2008worst}.

\subsection{Comparison with Mollison and Anderson \cite{mollison2013bringing}}
Mollison and Anderson \cite{mollison2013bringing} provide a library which performs a Global Earliest Deadline First (G-EDF) schedule on behalf of the OS. The library spawns worker threads on cores, similar to our approach, but it does not reserve one core for a scheduling thread as we do. Instead, they rely on a global queue, which is shared among all worker threads, and on test-and-set primitives to ensure mutually exclusive accesses. The code provided by Mollison and Anderson only includes an x86 version. Therefore, we adapt the architecture-dependent part of the code to run this experiment on our ARM-based platform. We also adapt their method to measure time to match ours, thus ensuring a fair comparison. 

Since Mollison and Anderson's library targets homogeneous multi-core architecture, we successively use 2 and 3 big cores to execute RT tasks. As \LibName{} runs a separate scheduler thread, we map this thread on the remaining big core. 

We use the task set generator based on the Dirichlet-Rescale (DRS) algorithm \cite{griffin2020generating}, which allows us to uniformly generate task sets with varying utilisation. We vary the number of tasks in the range $[20;120]$. For each number of cores, and for each number of tasks, we generate $5$ task sets, with utilisation varying in the range $[0.2;2]$. This results in $1360$ different task sets. The code related to each task is the same one as in \cite{mollison2013bringing}, which is a simple function that iterates to reach a pre-defined WCET.

\begin{figure*}[ht]
    \centering
    \begin{subfigure}{0.74\textwidth}
    \centering
        \includegraphics[width=1\textwidth]{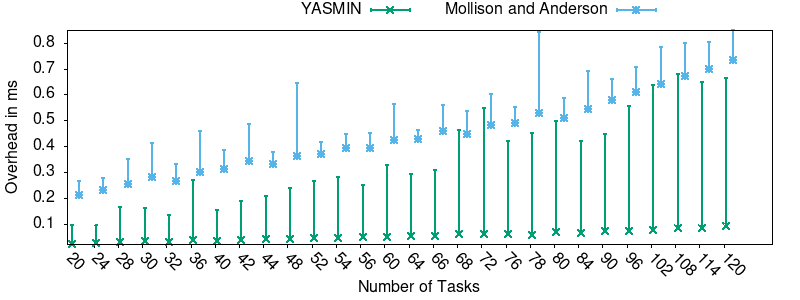}
        \caption{Average and maximum scheduling overhead by number of tasks}
        \label{fig:exp:mollison:tasks}
    \end{subfigure} 
    \begin{subfigure}{0.25\textwidth}
        \includegraphics[width=1\textwidth]{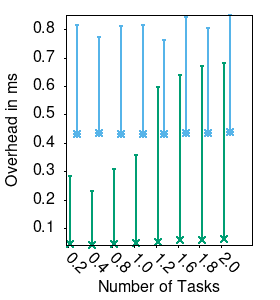}
        \caption{Average and maximum scheduling overhead by utilisation}
        \label{fig:exp:mollison:util}
    \end{subfigure} 
    \caption{Comparison of the overhead between \LibName{} and \cite{mollison2013bringing}}
    \label{fig:exp:mollison}
\end{figure*}

Figure~\ref{fig:exp:mollison} shows the evolution of the overhead depending on the amount of tasks and the total utilisation. 
\LibName{} demonstrates lesser overhead and better scalability in the number of tasks on average. However, the worst-case overhead observed with our library is a bit too high compared to the average, showing us how to drive further improvements. 

\subsection{Latency estimation comparison}

\textit{Cyclictest}\footnote{https://wiki.linuxfoundation.org/realtime/documentation/howto/tools/cyclictest/start} is a popular program used to accurately and repeatedly measures the response latency of a sporadic task activation. This program is available for \textit{Linux+PREEMPT\_RT} patch \textit{Linux+SCHED\_DEADLINE} and \textit{Litmus\^{}RT}. We adapt \textit{cyclictest} to run under \LibName{} management.

On the Odroid-XU4 board we switch between the different kernel \textit{Linux+PREEMPT\_RT} patch, and \textit{Litmus\^{}RT} to perform each run of \textit{cyclictest}. Unfortunately, we cannot include \textit{SCHED\_DEADLINE} scheduler class in our comparison as it is not available for our platform, which increases the motivation for creating \LibName{}. Similarly, \textit{Litmus\^{}RT} offers several scheduling policies, however \textit{cyclictest} fails to execute with some of them, and are therefore not included in this experiment.

We invoke \textit{cyclictest} with the same condition: \textit{-t 6 -d 0 -i 10000 -m  -l 10000}, i.e.~we want 6 threads woken up 10000 times at the same time with a 10ms period, and with locked memory. We restrict ourselves to 6 threads as our middleware library needs a $7^{th}$ thread for scheduling and we leave $1$ core available to the OS.

To generate interfering load on the platform we use the tool \textit{stress-ng}\footnote{https://wiki.ubuntu.com/Kernel/Reference/stress-ng}, which we configure to stress the scheduler and the computing cores. \textit{Stress-ng} is invoked with the following parameters: \textit{-C 8 -c 8 -T 8 -y 8}, which roughly means that 8 threads are spawned per stressor, i.e. cache trashing, computation, timer events, sched\_yield calls. For more details see the \textit{stress-ng} documentation.

Table \ref{tbl:exp:latency} displays the latencies we observed in the different configurations. The first column shows the kernel type and version used to gather the measurements. The second column displays the version of cyclictest used: \textit{\LibName{}} stands for our adapted version using our library, \textit{RTapps} stands for the common version shipped with the \textit{PREEMPT\_RT} patch set and \textit{litmus+XX} stands for the version shipped by \textit{Litmus\^{}RT} where $XX$ is the OS set up scheduler. The third column of Table \ref{tbl:exp:latency} shows the minimum, maximum and average latency observed across the 6 threads. 

On the Linux kernel with \textit{PREEMPT\_RT} the observed latency using \LibName{} is similar to the initial \textit{cyclictest} version, though slightly higher, due to our library overhead. When running on \textit{Litmus\^{}RT}, we observe a higher overhead of our library compared to other versions. However, the benefit of \textit{Litmus\^{}RT} comes at the price of no support for complex COTS heterogeneous platforms.

\begin{table}[ht]
    \centering
    \caption{Latency comparison between \LibName{}, Linux+PREEMPT\_RT and Litmus\^{}RT}
    \label{tbl:exp:latency}
    \begin{tabular}{|c|l|c|}
        \hline 
        OS & \textit{Cyclictest}  & Latency in $\mu s$\\
            &    version         & $<min,max,avg>$ \\
        \hline
        \makecell[l]{\textit{Linux}\\\textit{+PREEMPT\_RT}}     & \LibName{} & 90, 1481, 500 \\
        4.14.134-rt63    & RTapps & 176, 1550, 463 \\
        \hline
        \multirow{2}{*}{\textit{Litmus\^{}RT}}
            & \LibName{} & 67, 318, 170\\
            & RTapps & 33, 222, 74\\
        \multirow{2}{*}{4.9.30-litmus}    & litmus+GSN-EDF & 35, 247, 84\\
            & litmus+P-RES & 988, 1206, 1027\\
        \hline
    \end{tabular}
\end{table}
\section{\label{sec:use-case}An Unmanned Aerial Vehicule}

An industrial partner provided us with a use-case involving an Unmanned Aerial Vehicule (UAV) performing object detection on images. The goal of the use-case is to detect life boats on sea to call upon a rescue team and save lives.
Combining real-time requirements, rapid prototyping, scheduling exploration and task implementation exploration are the leitmotiv brought by our industrial use-case for the creation of our middleware. 
As compared to previous approaches \cite{mollison2013bringing, saranya2014implementation}, \LibName~ enables various scheduling policies, task models and platforms with a simple API.

\begin{figure}[ht]
\centering
    \begin{subfigure}[b]{0.39\textwidth}
        \includegraphics[width=1\textwidth]{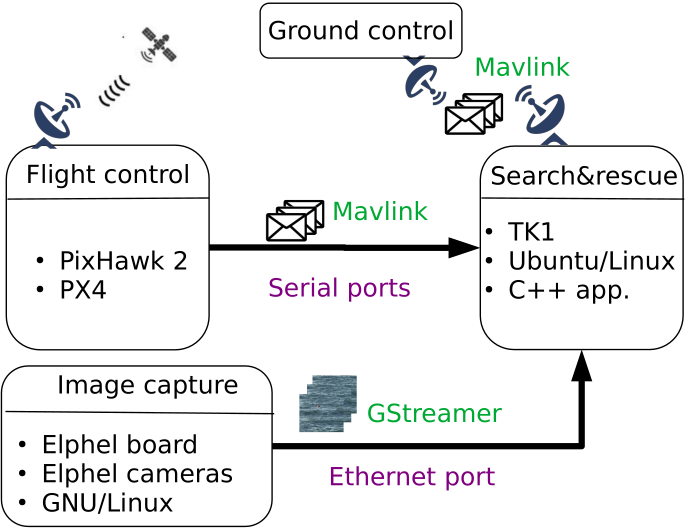}
        \caption{Overview of the system including hardware, operating environment, and software}
        \label{fig:uc:system}
    \end{subfigure}
    \begin{subfigure}[b]{0.48\textwidth}
        \includegraphics[width=1\textwidth]{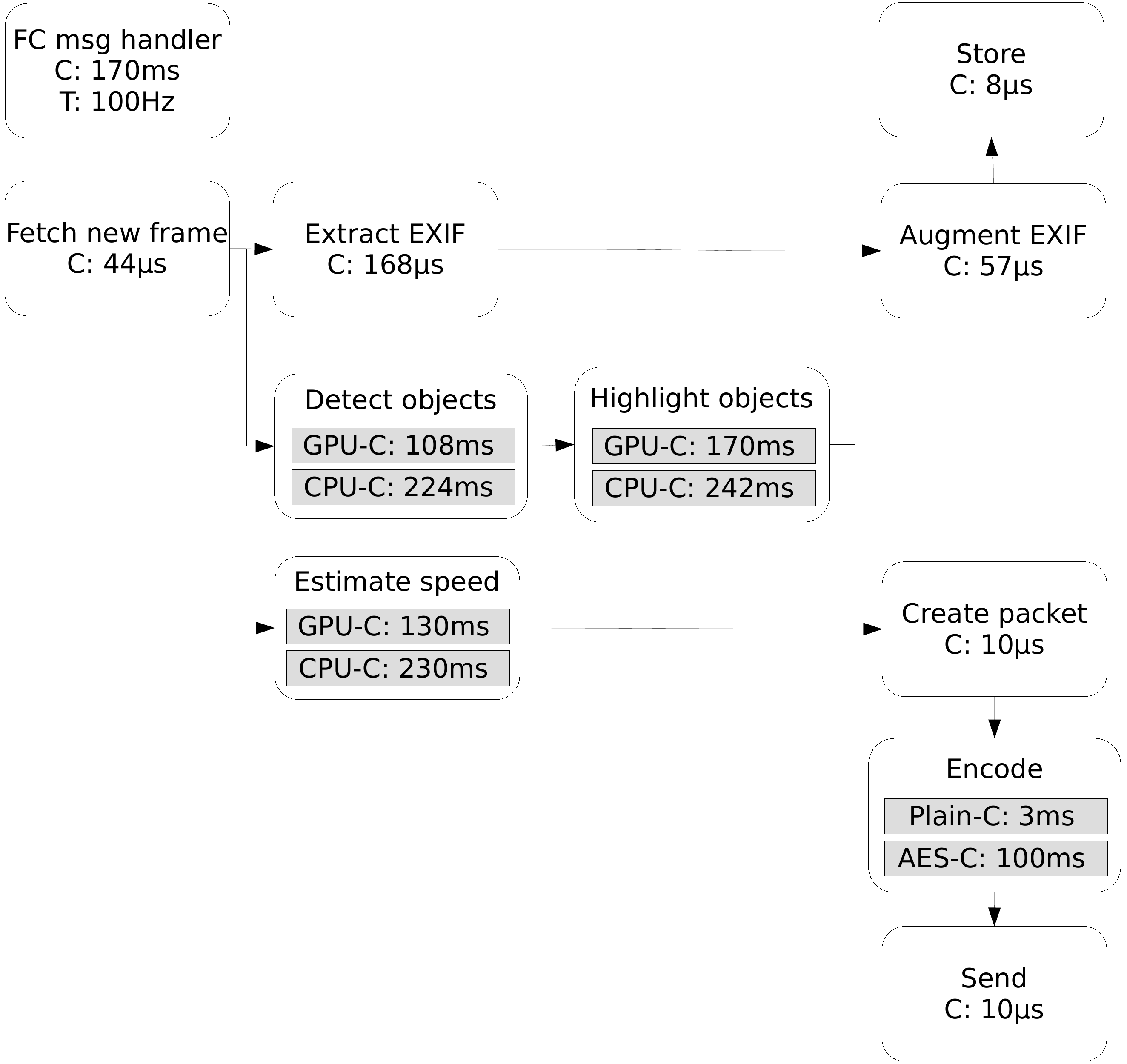}
        \caption{Simplified view of the SAR application tasks}
        \label{fig:uc:tasks}
    \end{subfigure}
    \caption{Drone use-case overview}
\end{figure}

The UAV under study is a fixed-wing drone. The application scenario is a Search \& Rescue (SAR) mission where the drone flies above the sea and sends an alarm to a ground station when it detects life boats. Figure \ref{fig:uc:system} provides a graphical sketch of the system. The drone embeds multiple computing platforms that can be split in three parts: flight control, image capture, and mission-specific payload application (here SAR).

\textbf{Flight Control: }
To fly in total autonomy the drone uses a GPS-based autopilot open-source software stack, called PX4\footnote{https://px4.io/ -- https://pixhawk.org/}, that pilots the drone following a pre-loaded mission. It runs on a PixHawk 2 platform\footnotemark[4] (single-core Cortex M4F with 256 KB RAM).

\textbf{Image Capture: }
To capture images an Elphel\footnote{https://www.elphel.com/ -- https://gstreamer.freedesktop.org/} board with a camera is mounted below the drone. The Elphel board runs GNU/Linux; captured frames are streamed using standard GStreamer\footnotemark[5] libraries. 


\textbf{Search \& Rescue Payload Application: }
The SAR application runs on a Toradex Apalis TK1\footnote{https://developer.toradex.com/products/apalis-tk1 -- https://ubuntu.com/} Computer-on-Module
hardware platform, which provides a quad-core ARM Cortex-A15 CPU, 2~GB of DDR3 RAM,
and 16~GB of non-volatile storage. It also features an NVIDIA Kepler GPU with 192 cores. 
The GPU device can be exploited to accelerate image processing tasks.
The board runs a modified Ubuntu/Linux\footnotemark[6], which includes NVIDIA proprietary drivers for the Kepler GPU. This precludes the use of both a Real-Time Operating System (RTOS) and the RT-patch set for Linux as neither of them supports this hardware platform. 

The original SAR application code, as provided by our industrial partner, has mostly been developed in C++, with an object detection function in CUDA. 
It receives Mavlink-encoded\footnote{https://mavlink.io/en/}  messages from Flight Control through a serial port on the board. Among others, these messages provide time synchronisation, update GPS coordinates, and enable/disable the payload application. 
The latter feature allows us to save energy by not running the SAR application while navigating to and from the mission area.

The SAR application also receives frames from the Image Capture through its ethernet port. 
Upon reception of a \textit{toggle image capture} message from the Flight Control, a GStreamer pipeline is activated. It downloads a new frame at a fixed frame rate. This frame is stored in a queue until it is processed by the detection algorithm, which is likewise activated/deactivated by the same message. Due to the low speed of the drone, there is no need for a high frame rate. The frame rate is set at 2 frames per second (fps).
Upon detecting life boats a message is sent to the Ground Control, including the number of boats, their corresponding GPS location, and the image itself for manual validation. 

Figure \ref{fig:uc:tasks} shows a simplified view of the tasks within the SAR. There are $2$ independent tasks, where one is a graph with multiple nodes. The periodicity of each root node is presented on the figure as well as their WCET. Also, $4$ tasks have multiple versions where $3$ of them (Detect objects, Highlight objects, and Estimate speed) deal with images with either a CUDA or a CPU only implementation, and $1$ task (Encode) that has $2$ implementations to either not encode the data $Plain$, or use the $AES$ algorithm. The later allows $2$ modes of execution: a normal mode, and a secure mode which is activated when boats are detected in the frame.

\begin{figure*}[ht]
    \centering
    \includegraphics[width=0.9\textwidth]{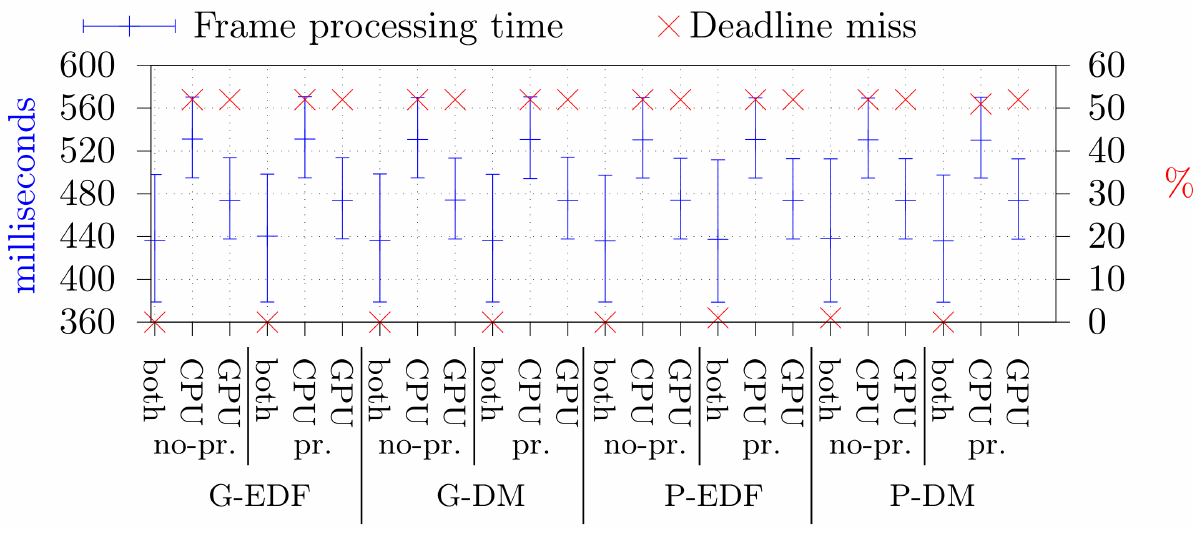}
    \caption{Scheduling exploration for the drone use-case}
    \label{fig:uc:result}
\end{figure*}

We crafted a simple mission to simulate an environment for the SAR application which has been compiled with \LibName{}. Figure \ref{fig:uc:result} shows, for different scheduling configurations, the measured time to \textit{process a frame} and the deadline miss ratio. For each configuration we forced the scheduler to use only CPU version of tasks, or only GPU version, or we allowed both versions and left the scheduler decides which one to execute.

As expected, the average time to process a frame is shorter for each scheduling strategy including the GPU. Both the CPU only, and GPU only configurations exhibit deadline misses in the same proportion. Even, if using the GPU only shorten the frame processing time, it is still too high and some frames miss their deadlines, but the other task to fetch messages from the Flight Control also misses a lot of deadlines due to CPU being overbooked. The only configurations decreasing deadline misses include both CPU and GPU versions, with automatic selection by the scheduler in \LibName{}, hence demonstrating the benefit from using multiple versions.

In the overall, all scheduling strategies (G-EDF, G-DM, P-EDF, P-DM) display the same overhead and deadline misses. Looking closer to the number, the partitioned strategies suffer from a tiny higher processing time per frame (6ms for the max values), which is due to the loss of the flexibility of a partitioned approach versus a global one. This leads to 1 deadline miss for P-EDF-both and P-DM-both for a Flight Control message.
\section{\label{sec:rw}Related Work}

Several resource allocation software have been proposed in the past in the form of RTOSes \cite{calandrino2006litmus,erika, gracioli2013implementation}, kernel patches \cite{balsini2014sched}, or as hypervisors \cite{martins2020bao}. These works enhance real-time scheduling, but they are, to some different degrees, not portable, easy to maintain or customisable as a user-space library or a middleware is. They also hardly support heterogeneous platforms, and are mostly limited to micro-controller.

Mollison and Anderson \cite{mollison2013bringing} created a library to schedule a set of task from the user-space. The library is intended to be used on a RTOS (Linux + PREEMPT\_RT in experiment). Target applications includes sporadic task sets scheduled on multiple cores grouped in cluster (C-EDF). Scheduling strategy allows dynamic priority (EDF), preemption and migration. Among scheduling capabilities and other issues, the library provides locking mechanism with priority inversion (short wait time: spin-lock, long wait time: context switch), synchronization protocols for critical sections, long system calls handling mechanism not blocking the overall system. The authors also provide an empirical evaluation of the different overhead induced by the library based on measurements. Similarly to them we abstract the schedule capabilities of the kernel within worker threads, or virtual processors, each mapped to a specific core. These worker threads are responsible to execute the real-time tasks and guarantee timing constraints. As opposed to them we do not allow job migration which makes our library more simple and with less overhead at the cost of specific scheduling strategies. In comparison to them, we advocate the reservation of a specific core for all non-RT tasks (mostly system tasks) and interrupt handlers, which allows us to provide the same guarantees in a simpler way with not only RTOS and COTS OS. Lastly, the implementation provided by the authors extensively use dynamic allocation which leads to hazard when estimating the WCET of this library, thus losing confidence in the reported overhead.

ExSched \cite{aasberg2012exsched} is a framework to allow scheduling from user space. It is composed of two parts: \begin{enumerate}
	\item A user space library providing a minimal API to final program to set user components parameters and controlling the beginning or end of a schedule
	\item A kernel space module which acts as a proxy between user-space API call and kernel scheduling primitives.
\end{enumerate} 
While the authors claim their method to be OS
independent, the ExSched library requires a Linux kernel module to be loaded, which strongly link their user-space API to the Linux kernel. However, as opposed to \LibName{}, they allow preemption and migration. An extension to mixed-critical tasks has been proposed in \cite{gupta2017extending}.

The ShedISA framework is introduced in \cite{saranya2014implementation} to enforce real-time constraints on COTS platform. This framework comes as an extension of the Linux kernel by providing a new scheduling class called \textit{SCHED\_IS}, which extends \textit{SCHED\_DEADLINE} but with a higher priority within the kernel. It heavily uses processor shielding by splitting cores in 3 groups, Linux cores: to execute system tasks, Service core: to execute the scheduler, RT cores: to execute SchedISA RT tasks. \ignore{They describe all the process to create a task which first start by executing non-rt code fragment, and then request a change of scheduler to be migrated to RT cores.} They only support P-EDF algorithm. But, a wide study of all induced overhead is presented.

SF3P (Scheduling Framework For Fast Prototyping) \cite{gomez2014sf3p} is a framework to explore the design of a hierarchical composition of real-time schedulers. This type of scheduler can be represented as a tree of schedulers where the next task to schedule is decided by going from the root of the tree to the leaf. Walking through the tree following each stage scheduler decisions at the end will effectively schedule a workload on a core. The framework allows to quickly build this scheduler tree in order to test its viability. However, the proposed framework is not meant for deployment and does not provide any timing guarantee as its purpose is for design space exploration only.

Serra et al. \cite{serra2020architecture} propose a complete middleware framework to enhance user experience regarding scheduling strategies from the Linux user-space. From the user point of view, it facilitates the setup of existing scheduling strategies by hiding required invocations to syscalls (pthread\_* API) that configure the environment. The framework is composed of a set of plugins, dynamically linked library, a daemon running in privileged mode (root), a user library linked to the final application. Each plugin corresponds to a specific scheduling policy that will interact with the kernel using the current available kernel API. These plugins are loaded by the daemon to apply the user configuration on threads in order to achieve the desired schedule. While the overall structure of the framework is kernel version independent, the related plugins are not, which make retro-compatibility and future maintenance complicated. Each task is considered as a thread which, in a very large system, is not applicable. In addition, only task addition overhead is presented, leaving other overheads unknown, e.g. interference with the daemon thread.

Similarly, Chishiro \cite{chishiro2016rt} proposes a middleware that sets the priority on threads in user-space to influence the scheduling decisions made by the kernel. This middleware, called RT-Seed, targets real-time trading systems with a parallel-extended imprecise computation task model executing a partitioned semi fixed priority scheduling algorithm. The type of targeted system include homogeneous processors, and the task model is unsuitable for embedded systems.

Singhal et al. \cite{singhal2014extended} propose to add a module to the kernel to add a scheduling level. This module will receive tasks (here tasks are processes) from the user-space, and then performs a schedulability test which, upon success, will compute the scheduling parameters. The upper scheduling level offered by the system will then schedule the processes according to their parameters. In order to achieve the desired scheduling policy, the module keeps track on which tasks have been added with which parameters to update them if necessary. A module is quite dependent to the current kernel version, at least to the major revision. No overhead analysis is presented in the paper.

Slite \cite{gadepalli2020slite} allows to control the scheduling of an entire system at the user-level. It augments the Composite OS \cite{parmer2008predictable} with a direct mapping between kernel and user threads. Both levels exchange messages to maintain the coherence of active threads, this allows to account for interrupt threads placed by the kernel in the scheduling policy\ignore{ enforced at user level managing user threads}. They support partitioned (non-)preemptive fixed priorities and EDF where each core has it own scheduling thread.

Finally, Bristot de Oliveira et al. \cite{de2020demystifying} shows how to account for scheduling overhead within the Linux kernel. Thus helping to identify where overhead occurs withing schedule deployment.

A major difference with all the listed frameworks and libraries is their focus on on-line scheduling strategy while we also enable pre-computed off-line scheduling strategies. On the task model they only focus on independent tasks while we allow graph-based task models. Finally, we are not aware of any previous work dealing with heterogeneous architectures, and multiple versions per tasks with a runtime offering automatic selection based on different criteria.
\section{\label{sec:concl}Conclusions}

We have presented \LibName{}, a middleware that performs the scheduling of an end-user application on behalf of the OS in user space.
Our middleware significantly simplifies design space exploration with respect to scheduling and mapping by disentangling scheduling and mapping decisions from functional application code.

\LibName{} is, to the best of our knowledge, the first library that embraces heterogeneity on embedded COTS platforms, among others through support for multi-version tasks. The major features of \LibName~ includes the possibility to easily switch between scheduling policies, to deploy applications using scheduling policies not available at the OS level without the need to adapt the OS code nor the end user application. 
We show that the overhead and latency induced by \LibName~ is on comparable with state of the art task management systems. We also showed the applicability and benefit of using \LibName~ and multi-version tasks on an industrial use-case with a Search \& Rescue drone.

As future work we plan to improve the management of real-time tasks with arbitrary activation patterns by using recurring servers, e.g.~\cite{ghazalie1995aperiodic}. 
This would increase our support for real-time applications. We also plan to improve the support for heterogeneous platforms, by adding a mechanism to provide asynchronous usage of hardware accelerators, and support for FPGA such as \cite{gracioli2019designing}. Finally, we plan to enable delay tokens mechanism, thus relaxing the acyclic constraint in graph-based task model.

\bibliography{biblio}


\begin{thebibliography}{39}


\ifx \showCODEN    \undefined \def \showCODEN     #1{\unskip}     \fi
\ifx \showDOI      \undefined \def \showDOI       #1{#1}\fi
\ifx \showISBNx    \undefined \def \showISBNx     #1{\unskip}     \fi
\ifx \showISBNxiii \undefined \def \showISBNxiii  #1{\unskip}     \fi
\ifx \showISSN     \undefined \def \showISSN      #1{\unskip}     \fi
\ifx \showLCCN     \undefined \def \showLCCN      #1{\unskip}     \fi
\ifx \shownote     \undefined \def \shownote      #1{#1}          \fi
\ifx \showarticletitle \undefined \def \showarticletitle #1{#1}   \fi
\ifx \showURL      \undefined \def \showURL       {\relax}        \fi
\providecommand\bibfield[2]{#2}
\providecommand\bibinfo[2]{#2}
\providecommand\natexlab[1]{#1}
\providecommand\showeprint[2][]{arXiv:#2}

\bibitem[\protect\citeauthoryear{}{}{  a}]%
        {rouxel2020prego}
\bibfield{author}{\bibinfo{person}{}.} \bibinfo{year}{--}\natexlab{a}.
\newblock \showarticletitle{{Removed for double-blind review.}}. In
  \bibinfo{booktitle}{\emph{--}}. \bibinfo{publisher}{--}.
\newblock


\bibitem[\protect\citeauthoryear{}{}{  b}]%
        {roeder2020towards}
\bibfield{author}{\bibinfo{person}{}.} \bibinfo{year}{--}\natexlab{b}.
\newblock \showarticletitle{{Removed for double-blind review}}. In
  \bibinfo{booktitle}{\emph{--}}.
\newblock


\bibitem[\protect\citeauthoryear{}{}{[n.d.]}]%
        {splorer}
\bibfield{author}{\bibinfo{person}{}.} \bibinfo{year}{[n.d.]}\natexlab{}.
\newblock \bibinfo{booktitle}{\emph{{Removed for double-blind review}}}.
\newblock
\urldef\tempurl%
\url{gitLab Repository}
\showURL{%
\tempurl}


\bibitem[\protect\citeauthoryear{??}{eri}{2018}]%
        {erika}
 \bibinfo{year}{2018}\natexlab{}.
\newblock \bibinfo{title}{Erika Enterprise
  \url{{http://erika.tuxfamily.org/drupal}}}.
\newblock
\newblock


\bibitem[\protect\citeauthoryear{{AA}sberg, Nolte, Kato, and
  Rajkumar}{{AA}sberg et~al\mbox{.}}{2012}]%
        {aasberg2012exsched}
\bibfield{author}{\bibinfo{person}{Mikael {AA}sberg}, \bibinfo{person}{Thomas
  Nolte}, \bibinfo{person}{Shinpei Kato}, {and} \bibinfo{person}{Ragunathan
  Rajkumar}.} \bibinfo{year}{2012}\natexlab{}.
\newblock \showarticletitle{{Exsched: An External CPU Scheduler Framework for
  Real-Time Systems}}. \bibinfo{pages}{240--249}.
\newblock


\bibitem[\protect\citeauthoryear{Akesson, Nasri, Nelissen, Altmeyer, and
  Davis}{Akesson et~al\mbox{.}}{2020}]%
        {akesson2020empirical}
\bibfield{author}{\bibinfo{person}{Benny Akesson}, \bibinfo{person}{Mitra
  Nasri}, \bibinfo{person}{Geoffrey Nelissen}, \bibinfo{person}{Sebastian
  Altmeyer}, {and} \bibinfo{person}{Robert~Ian Davis}.}
  \bibinfo{year}{2020}\natexlab{}.
\newblock \showarticletitle{An empirical survey-based study into industry
  practice in real-time systems}. In \bibinfo{booktitle}{\emph{2020 IEEE
  Real-Time Systems Symposium (Proceedings)}}.
\newblock


\bibitem[\protect\citeauthoryear{Balsini}{Balsini}{2014}]%
        {balsini2014sched}
\bibfield{author}{\bibinfo{person}{Alessio Balsini}.}
  \bibinfo{year}{2014}\natexlab{}.
\newblock \showarticletitle{{Adaptive Scheduling Parameters Manager for
  SCHED\_DEADLINE}}. In \bibinfo{booktitle}{\emph{Workshop on Real-Time
  Scheduling in the Linux Kernel}}.
\newblock


\bibitem[\protect\citeauthoryear{Brosky and Rotolo}{Brosky and Rotolo}{2003}]%
        {brosky2003shielded}
\bibfield{author}{\bibinfo{person}{Steve Brosky} {and} \bibinfo{person}{Steve
  Rotolo}.} \bibinfo{year}{2003}\natexlab{}.
\newblock \showarticletitle{{Shielded processors: Guaranteeing sub-millisecond
  response in standard Linux}}. In \bibinfo{booktitle}{\emph{International
  Parallel and Distributed Processing Symposium (IPDPS)}}.
  \bibinfo{pages}{9--pp}.
\newblock


\bibitem[\protect\citeauthoryear{Calandrino, Leontyev, Block, Devi, and
  Anderson}{Calandrino et~al\mbox{.}}{2006}]%
        {calandrino2006litmus}
\bibfield{author}{\bibinfo{person}{John~M Calandrino},
  \bibinfo{person}{Hennadiy Leontyev}, \bibinfo{person}{Aaron Block},
  \bibinfo{person}{UmaMaheswari~C Devi}, {and} \bibinfo{person}{James~H
  Anderson}.} \bibinfo{year}{2006}\natexlab{}.
\newblock \showarticletitle{{Litmus\^{}rt: A testbed for empirically comparing
  real-time multiprocessor schedulers}}. In \bibinfo{booktitle}{\emph{Real-Time
  Systems Symposium (RTSS)}}. \bibinfo{pages}{111--126}.
\newblock


\bibitem[\protect\citeauthoryear{Calvaresi, Marinoni, Sturm, Schumacher, and
  Buttazzo}{Calvaresi et~al\mbox{.}}{2017}]%
        {calvaresi2017challenge}
\bibfield{author}{\bibinfo{person}{Davide Calvaresi}, \bibinfo{person}{Mauro
  Marinoni}, \bibinfo{person}{Arnon Sturm}, \bibinfo{person}{Michael
  Schumacher}, {and} \bibinfo{person}{Giorgio Buttazzo}.}
  \bibinfo{year}{2017}\natexlab{}.
\newblock \showarticletitle{{The challenge of real-time multi-agent systems for
  enabling IoT and CPS}}. In \bibinfo{booktitle}{\emph{International Conference
  on Web Intelligence (WI)}}. \bibinfo{pages}{356--364}.
\newblock


\bibitem[\protect\citeauthoryear{Chishiro}{Chishiro}{2016}]%
        {chishiro2016rt}
\bibfield{author}{\bibinfo{person}{Hiroyuki Chishiro}.}
  \bibinfo{year}{2016}\natexlab{}.
\newblock \showarticletitle{{Rt-seed: Real-time middleware for
  semi-fixed-priority scheduling}}. In \bibinfo{booktitle}{\emph{International
  Symposium on Real-Time Distributed Computing (ISORC)}}.
  \bibinfo{pages}{124--133}.
\newblock


\bibitem[\protect\citeauthoryear{de~Oliveira, Casini, de~Oliveira, and
  Cucinotta}{de~Oliveira et~al\mbox{.}}{2020}]%
        {de2020demystifying}
\bibfield{author}{\bibinfo{person}{Daniel~Bristot de Oliveira},
  \bibinfo{person}{Daniel Casini}, \bibinfo{person}{R{\^o}mulo~Silva de
  Oliveira}, {and} \bibinfo{person}{Tommaso Cucinotta}.}
  \bibinfo{year}{2020}\natexlab{}.
\newblock \showarticletitle{{Demystifying the Real-Time Linux Scheduling
  Latency}}. In \bibinfo{booktitle}{\emph{32nd Euromicro Conference on
  Real-Time Systems (ECRTS 2020)}}. Schloss Dagstuhl-Leibniz-Zentrum f{\"u}r
  Informatik.
\newblock


\bibitem[\protect\citeauthoryear{Ferdinand and Heckmann}{Ferdinand and
  Heckmann}{2004}]%
        {ferdinand2004ait}
\bibfield{author}{\bibinfo{person}{Christian Ferdinand} {and}
  \bibinfo{person}{Reinhold Heckmann}.} \bibinfo{year}{2004}\natexlab{}.
\newblock \showarticletitle{{aiT: Worst-case execution time prediction by
  static program analysis}}.
\newblock In \bibinfo{booktitle}{\emph{Building the Information Society}}.
  \bibinfo{publisher}{Springer}, \bibinfo{pages}{377--383}.
\newblock


\bibitem[\protect\citeauthoryear{Gadepalli, Pan, and Parmer}{Gadepalli
  et~al\mbox{.}}{2020}]%
        {gadepalli2020slite}
\bibfield{author}{\bibinfo{person}{Phani~Kishore Gadepalli},
  \bibinfo{person}{Runyu Pan}, {and} \bibinfo{person}{Gabriel Parmer}.}
  \bibinfo{year}{2020}\natexlab{}.
\newblock \showarticletitle{{Slite: OS Support for Near Zero-Cost, Configurable
  Scheduling}}. In \bibinfo{booktitle}{\emph{Real-Time and Embedded Technology
  and Applications Symposium (RTAS)}}. \bibinfo{pages}{160--173}.
\newblock


\bibitem[\protect\citeauthoryear{Ghazalie and Baker}{Ghazalie and
  Baker}{1995}]%
        {ghazalie1995aperiodic}
\bibfield{author}{\bibinfo{person}{Teguh~M Ghazalie} {and}
  \bibinfo{person}{Theodore~P. Baker}.} \bibinfo{year}{1995}\natexlab{}.
\newblock \showarticletitle{{Aperiodic servers in a deadline scheduling
  environment}}.
\newblock \bibinfo{journal}{\emph{Real-Time Systems}} \bibinfo{volume}{9},
  \bibinfo{number}{1} (\bibinfo{year}{1995}), \bibinfo{pages}{31--67}.
\newblock


\bibitem[\protect\citeauthoryear{Gomez, Schor, Kumar, and Thiele}{Gomez
  et~al\mbox{.}}{2014}]%
        {gomez2014sf3p}
\bibfield{author}{\bibinfo{person}{Andres Gomez}, \bibinfo{person}{Lars Schor},
  \bibinfo{person}{Pratyush Kumar}, {and} \bibinfo{person}{Lothar Thiele}.}
  \bibinfo{year}{2014}\natexlab{}.
\newblock \showarticletitle{{Sf3p: A framework to explore and prototype
  hierarchical compositions of real-time schedulers}}. In
  \bibinfo{booktitle}{\emph{International Symposium on Rapid System Prototyping
  (RSP)}}. \bibinfo{pages}{2--8}.
\newblock


\bibitem[\protect\citeauthoryear{Goossens, Poczekajlo, Paolillo, and
  Rodriguez}{Goossens et~al\mbox{.}}{2019}]%
        {goossens2019acceptor}
\bibfield{author}{\bibinfo{person}{Jo{\"e}l Goossens}, \bibinfo{person}{Xavier
  Poczekajlo}, \bibinfo{person}{Antonio Paolillo}, {and} \bibinfo{person}{Paul
  Rodriguez}.} \bibinfo{year}{2019}\natexlab{}.
\newblock \showarticletitle{{ACCEPTOR: a model and a protocol for real-time
  multi-mode applications on reconfigurable heterogeneous platforms}}. In
  \bibinfo{booktitle}{\emph{Proceedings of the 27th International Conference on
  Real-Time Networks and Systems}}. \bibinfo{pages}{209--219}.
\newblock


\bibitem[\protect\citeauthoryear{Gracioli, Fr{\"o}hlich, Pellizzoni, and
  Fischmeister}{Gracioli et~al\mbox{.}}{2013}]%
        {gracioli2013implementation}
\bibfield{author}{\bibinfo{person}{Giovani Gracioli},
  \bibinfo{person}{Ant{\^o}nio~Augusto Fr{\"o}hlich}, \bibinfo{person}{Rodolfo
  Pellizzoni}, {and} \bibinfo{person}{Sebastian Fischmeister}.}
  \bibinfo{year}{2013}\natexlab{}.
\newblock \showarticletitle{{Implementation and evaluation of global and
  partitioned scheduling in a real-time OS}}.
\newblock \bibinfo{journal}{\emph{Real-Time Systems}} \bibinfo{volume}{49},
  \bibinfo{number}{6} (\bibinfo{year}{2013}), \bibinfo{pages}{669--714}.
\newblock


\bibitem[\protect\citeauthoryear{Gracioli, Tabish, Mancuso, Mirosanlou,
  Pellizzoni, and Caccamo}{Gracioli et~al\mbox{.}}{2019}]%
        {gracioli2019designing}
\bibfield{author}{\bibinfo{person}{Giovani Gracioli}, \bibinfo{person}{Rohan
  Tabish}, \bibinfo{person}{Renato Mancuso}, \bibinfo{person}{Reza Mirosanlou},
  \bibinfo{person}{Rodolfo Pellizzoni}, {and} \bibinfo{person}{Marco Caccamo}.}
  \bibinfo{year}{2019}\natexlab{}.
\newblock \showarticletitle{Designing mixed criticality applications on modern
  heterogeneous mpsoc platforms}. In \bibinfo{booktitle}{\emph{31st Euromicro
  Conference on Real-Time Systems (ECRTS 2019)}}. Schloss
  Dagstuhl-Leibniz-Zentrum fuer Informatik.
\newblock


\bibitem[\protect\citeauthoryear{Griffin, Bate, and Davis}{Griffin
  et~al\mbox{.}}{2020}]%
        {griffin2020generating}
\bibfield{author}{\bibinfo{person}{David~Jack Griffin},
  \bibinfo{person}{Iain~John Bate}, {and} \bibinfo{person}{Robert~Ian Davis}.}
  \bibinfo{year}{2020}\natexlab{}.
\newblock \showarticletitle{Generating Utilization Vectors for the Systematic
  Evaluation of Schedulability Tests}. In \bibinfo{booktitle}{\emph{2020 IEEE
  Real-Time Systems Symposium (proceedings)}}. York.
\newblock


\bibitem[\protect\citeauthoryear{Gupta, Luit, Heuvel, and Bril}{Gupta
  et~al\mbox{.}}{2017}]%
        {gupta2017extending}
\bibfield{author}{\bibinfo{person}{Tarun Gupta}, \bibinfo{person}{Erik~J Luit},
  \bibinfo{person}{Martijn~MHPVD Heuvel}, {and} \bibinfo{person}{Reinder~J
  Bril}.} \bibinfo{year}{2017}\natexlab{}.
\newblock \showarticletitle{{Extending ExSched with Mixed Criticality Support
  — An Experience Report}}. In \bibinfo{booktitle}{\emph{International
  Conference on Software Architecture Workshops (ICSAW)}}.
  \bibinfo{pages}{23--28}.
\newblock


\bibitem[\protect\citeauthoryear{HardKernel}{HardKernel}{2017}]%
        {odroidxu4}
\bibfield{author}{\bibinfo{person}{HardKernel}.}
  \bibinfo{year}{2017}\natexlab{}.
\newblock \bibinfo{booktitle}{\emph{{Odroid-XU4}, User ManualReal-Time Systems
  Symposium}}.
\newblock
\urldef\tempurl%
\url{https://magazine.odroid.com/odroid-xu4/}
\showURL{%
\tempurl}


\bibitem[\protect\citeauthoryear{Hardy, Rouxel, and Puaut}{Hardy
  et~al\mbox{.}}{2017}]%
        {hardy2017heptane}
\bibfield{author}{\bibinfo{person}{Damien Hardy}, \bibinfo{person}{Benjamin
  Rouxel}, {and} \bibinfo{person}{Isabelle Puaut}.}
  \bibinfo{year}{2017}\natexlab{}.
\newblock \showarticletitle{{The Heptane Static Worst-Case Execution Time
  Estimation Tool}}. In \bibinfo{booktitle}{\emph{Workshop on Worst-Case
  Execution Time Analysis (WCET)}}.
\newblock


\bibitem[\protect\citeauthoryear{Houssam-Eddine, Capodieci, Cavicchioli,
  Lipari, and Bertogna}{Houssam-Eddine et~al\mbox{.}}{2020}]%
        {houssam2020hpc}
\bibfield{author}{\bibinfo{person}{Zahaf Houssam-Eddine},
  \bibinfo{person}{Nicola Capodieci}, \bibinfo{person}{Roberto Cavicchioli},
  \bibinfo{person}{Giuseppe Lipari}, {and} \bibinfo{person}{Marko Bertogna}.}
  \bibinfo{year}{2020}\natexlab{}.
\newblock \showarticletitle{The HPC-DAG Task Model for Heterogeneous Real-Time
  Systems}.
\newblock \bibinfo{journal}{\emph{IEEE Trans. Comput.}} (\bibinfo{year}{2020}).
\newblock


\bibitem[\protect\citeauthoryear{Lee and Messerschmitt}{Lee and
  Messerschmitt}{1987}]%
        {lee1987static}
\bibfield{author}{\bibinfo{person}{Edward~Ashford Lee} {and}
  \bibinfo{person}{David~G Messerschmitt}.} \bibinfo{year}{1987}\natexlab{}.
\newblock \showarticletitle{{Static scheduling of synchronous data flow
  programs for digital signal processing}}.
\newblock \bibinfo{journal}{\emph{IEEE Trans. Comput.}} \bibinfo{volume}{100},
  \bibinfo{number}{1} (\bibinfo{year}{1987}), \bibinfo{pages}{24--35}.
\newblock


\bibitem[\protect\citeauthoryear{Martins, Tavares, Solieri, Bertogna, and
  Pinto}{Martins et~al\mbox{.}}{2020}]%
        {martins2020bao}
\bibfield{author}{\bibinfo{person}{Jos{\'e} Martins}, \bibinfo{person}{Adriano
  Tavares}, \bibinfo{person}{Marco Solieri}, \bibinfo{person}{Marko Bertogna},
  {and} \bibinfo{person}{Sandro Pinto}.} \bibinfo{year}{2020}\natexlab{}.
\newblock \showarticletitle{{Bao: A Lightweight Static Partitioning Hypervisor
  for Modern Multi-Core Embedded Systems}}. In
  \bibinfo{booktitle}{\emph{Workshop on Next Generation Real-Time Embedded
  Systems (NG-RES)}}.
\newblock


\bibitem[\protect\citeauthoryear{Mellor-Crummey and Scott}{Mellor-Crummey and
  Scott}{1991}]%
        {mellor1991algorithms}
\bibfield{author}{\bibinfo{person}{John~M Mellor-Crummey} {and}
  \bibinfo{person}{Michael~L Scott}.} \bibinfo{year}{1991}\natexlab{}.
\newblock \showarticletitle{{Algorithms for scalable synchronization on
  shared-memory multiprocessors}}.
\newblock \bibinfo{journal}{\emph{ACM Transactions on Computer Systems (TOCS)}}
  \bibinfo{volume}{9}, \bibinfo{number}{1} (\bibinfo{year}{1991}),
  \bibinfo{pages}{21--65}.
\newblock


\bibitem[\protect\citeauthoryear{Mollison and Anderson}{Mollison and
  Anderson}{2013}]%
        {mollison2013bringing}
\bibfield{author}{\bibinfo{person}{Malcolm~S Mollison} {and}
  \bibinfo{person}{James~H Anderson}.} \bibinfo{year}{2013}\natexlab{}.
\newblock \showarticletitle{{Bringing theory into practice: A userspace library
  for multicore real-time scheduling}}. In \bibinfo{booktitle}{\emph{Real-Time
  and Embedded Technology and Applications Symposium (RTAS)}}.
  \bibinfo{pages}{283--292}.
\newblock


\bibitem[\protect\citeauthoryear{Nvidia}{Nvidia}{[n.d.]}]%
        {nvidiatx2}
\bibfield{author}{\bibinfo{person}{Nvidia}.} \bibinfo{year}{[n.d.]}\natexlab{}.
\newblock \bibinfo{booktitle}{\emph{{Jetson TX2}, presentation}}.
\newblock
\urldef\tempurl%
\url{https://developer.nvidia.com/embedded/jetson-tx2}
\showURL{%
\tempurl}


\bibitem[\protect\citeauthoryear{Parmer and West}{Parmer and West}{2008}]%
        {parmer2008predictable}
\bibfield{author}{\bibinfo{person}{Gabriel Parmer} {and}
  \bibinfo{person}{Richard West}.} \bibinfo{year}{2008}\natexlab{}.
\newblock \showarticletitle{{Predictable interrupt management and scheduling in
  the Composite component-based system}}. In
  \bibinfo{booktitle}{\emph{Real-Time Systems Symposium}}.
  \bibinfo{pages}{232--243}.
\newblock


\bibitem[\protect\citeauthoryear{Rajkumar}{Rajkumar}{2012}]%
        {rajkumar2012synchronization}
\bibfield{author}{\bibinfo{person}{Ragunathan Rajkumar}.}
  \bibinfo{year}{2012}\natexlab{}.
\newblock \bibinfo{booktitle}{\emph{Synchronization in real-time systems: a
  priority inheritance approach}}. Vol.~\bibinfo{volume}{151}.
\newblock \bibinfo{publisher}{Springer Science \& Business Media}.
\newblock


\bibitem[\protect\citeauthoryear{Saranya and Hansdah}{Saranya and
  Hansdah}{2014}]%
        {saranya2014implementation}
\bibfield{author}{\bibinfo{person}{N Saranya} {and} \bibinfo{person}{RC
  Hansdah}.} \bibinfo{year}{2014}\natexlab{}.
\newblock \showarticletitle{{An implementation of partitioned scheduling scheme
  for hard real-time tasks in multicore linux with fair share for linux
  tasks}}. In \bibinfo{booktitle}{\emph{International Conference on Embedded
  and Real-Time Computing Systems and Applications (RTCSA)}}.
  \bibinfo{pages}{1--9}.
\newblock


\bibitem[\protect\citeauthoryear{Serra, Ara, Fara, and Cucinotta}{Serra
  et~al\mbox{.}}{2020}]%
        {serra2020architecture}
\bibfield{author}{\bibinfo{person}{Gabriele Serra}, \bibinfo{person}{Gabriele
  Ara}, \bibinfo{person}{Pietro Fara}, {and} \bibinfo{person}{Tommaso
  Cucinotta}.} \bibinfo{year}{2020}\natexlab{}.
\newblock \showarticletitle{{An Architecture for Declarative Real-Time
  Scheduling on Linux}}. In \bibinfo{booktitle}{\emph{International Symposium
  on Real-Time Distributed Computing (ISORC)}}. \bibinfo{pages}{20--28}.
\newblock


\bibitem[\protect\citeauthoryear{Singhal, Kumar, Ghintala, and Chakma}{Singhal
  et~al\mbox{.}}{2014}]%
        {singhal2014extended}
\bibfield{author}{\bibinfo{person}{Purnima Singhal}, \bibinfo{person}{Amit
  Kumar}, \bibinfo{person}{Upendra Ghintala}, {and} \bibinfo{person}{Kunal
  Chakma}.} \bibinfo{year}{2014}\natexlab{}.
\newblock \showarticletitle{{Extended Level Real time Scheduling Framework:
  Using a generalized non-real time platform}}. In
  \bibinfo{booktitle}{\emph{2014 International Conference on Advances in
  Computing, Communications and Informatics (ICACCI)}}.
  \bibinfo{pages}{1279--1284}.
\newblock


\bibitem[\protect\citeauthoryear{Software}{Software}{[n.d.]}]%
        {pclint}
\bibfield{author}{\bibinfo{person}{Gimpel Software}.}
  \bibinfo{year}{[n.d.]}\natexlab{}.
\newblock \bibinfo{booktitle}{\emph{PC-Lint Plus}}.
\newblock
\urldef\tempurl%
\url{https://www.gimpel.com/pclp.html}
\showURL{%
\tempurl}


\bibitem[\protect\citeauthoryear{{Tian}, {Shi}, {Wang}, {Zhu}, {Du}, {Su},
  {Sun}, and {Guizani}}{{Tian} et~al\mbox{.}}{2019}]%
        {tian2019real}
\bibfield{author}{\bibinfo{person}{Z. {Tian}}, \bibinfo{person}{W. {Shi}},
  \bibinfo{person}{Y. {Wang}}, \bibinfo{person}{C. {Zhu}}, \bibinfo{person}{X.
  {Du}}, \bibinfo{person}{S. {Su}}, \bibinfo{person}{Y. {Sun}}, {and}
  \bibinfo{person}{N. {Guizani}}.} \bibinfo{year}{2019}\natexlab{}.
\newblock \showarticletitle{{Real-Time Lateral Movement Detection Based on
  Evidence Reasoning Network for Edge Computing Environment}}.
\newblock \bibinfo{journal}{\emph{IEEE Transactions on Industrial Informatics}}
  \bibinfo{volume}{15}, \bibinfo{number}{7} (\bibinfo{year}{2019}),
  \bibinfo{pages}{4285--4294}.
\newblock


\bibitem[\protect\citeauthoryear{Toradex}{Toradex}{[n.d.]}]%
        {toradextk1}
\bibfield{author}{\bibinfo{person}{Toradex}.}
  \bibinfo{year}{[n.d.]}\natexlab{}.
\newblock \bibinfo{booktitle}{\emph{{Apalis TK1}, presentation}}.
\newblock
\urldef\tempurl%
\url{https://www.toradex.com/computer-on-modules/apalis-arm-family/nvidia-tegra-k1}
\showURL{%
\tempurl}


\bibitem[\protect\citeauthoryear{Vermesan, Friess, Guillemin, Gusmeroli,
  Sundmaeker, Bassi, Jubert, Mazura, Harrison, Eisenhauer,
  et~al\mbox{.}}{Vermesan et~al\mbox{.}}{2011}]%
        {vermesan2011internet}
\bibfield{author}{\bibinfo{person}{Ovidiu Vermesan}, \bibinfo{person}{Peter
  Friess}, \bibinfo{person}{Patrick Guillemin}, \bibinfo{person}{Sergio
  Gusmeroli}, \bibinfo{person}{Harald Sundmaeker}, \bibinfo{person}{Alessandro
  Bassi}, \bibinfo{person}{Ignacio~Soler Jubert}, \bibinfo{person}{Margaretha
  Mazura}, \bibinfo{person}{Mark Harrison}, \bibinfo{person}{Markus
  Eisenhauer}, {et~al\mbox{.}}} \bibinfo{year}{2011}\natexlab{}.
\newblock \showarticletitle{{Internet of Things Strategic Research Roadmap}}.
\newblock \bibinfo{journal}{\emph{Internet of Things -- Global Technological
  and Societal Trends}} (\bibinfo{year}{2011}), \bibinfo{pages}{9--52}.
\newblock


\bibitem[\protect\citeauthoryear{Wilhelm, Engblom, Ermedahl, Holsti, Thesing,
  Whalley, Bernat, Ferdinand, Heckmann, Mitra, et~al\mbox{.}}{Wilhelm
  et~al\mbox{.}}{2008}]%
        {wilhelm2008worst}
\bibfield{author}{\bibinfo{person}{Reinhard Wilhelm}, \bibinfo{person}{Jakob
  Engblom}, \bibinfo{person}{Andreas Ermedahl}, \bibinfo{person}{Niklas
  Holsti}, \bibinfo{person}{Stephan Thesing}, \bibinfo{person}{David Whalley},
  \bibinfo{person}{Guillem Bernat}, \bibinfo{person}{Christian Ferdinand},
  \bibinfo{person}{Reinhold Heckmann}, \bibinfo{person}{Tulika Mitra},
  {et~al\mbox{.}}} \bibinfo{year}{2008}\natexlab{}.
\newblock \showarticletitle{The worst-case execution-time problem—overview of
  methods and survey of tools}.
\newblock \bibinfo{journal}{\emph{ACM Transactions on Embedded Computing
  Systems (TECS)}} \bibinfo{volume}{7}, \bibinfo{number}{3}
  (\bibinfo{year}{2008}), \bibinfo{pages}{36}.
\newblock


\end{thebibliography}



\end{document}